\documentclass[%
 reprint,
 amsmath,amssymb,
 aps,
]{revtex4-2}

\usepackage{graphicx}
\usepackage{dcolumn}
\usepackage{bm}

\usepackage{color}
\newenvironment{psmallmatrix}
  {\left(\begin{smallmatrix}}
  {\end{smallmatrix}\right)}
\usepackage{physics}
\usepackage{multirow}
\usepackage{array}
\newcolumntype{P}[1]{>{\centering\arraybackslash}p{#1}}

\begin{document}

\preprint{APS/123-QED}

\title{Quantum Optics based Algorithm for Measuring the Similarity between Images}

\author{Vivek Mehta}
\author{Sonali Jana}%


\author{Utpal Roy}
\email{uroy@iitp.ac.in}
\affiliation{Indian Institute of Technology Patna, Bihta, Patna 801103, India}%


\date{\today}

\begin{abstract}
We report an algorithm, based on quantum optics formulation, where a coherent state is used as the elementary quantum resource for the image representation. We provide an architecture with constituent optical elements in linear order with respect to the image resolution. The obtained phase-distributed multimode coherent state is fed into an image retrieval scheme and we identify the appropriate laser intensity parameter for similarity measurement. The use of the principle of quantum superposition in the similarity measurement protocol enables us to encode multiple input images. We demonstrate the viability of the protocol through an objective quality assessment of images by adding consecutive layers of noises. The results are in good agreement with the expected outcome. The image distortion-sensitivity analysis of the metric establishes the further merit of the model. Our quantum algorithm has wider applicability also in supervised machine learning tasks.
\end{abstract}

\maketitle

\section{Introduction}
 \label{section 1}

Quantum mechanics offers a huge improvement in the domains of information processing, computation, and technology by demystifying many physical phenomena, which rely on its peculiar properties, such as the superposition of quantons, and quantum correlations. The applications also subsume quantum sensing \cite{degen2017quantum, bera2022quantum, giovannetti2006quantum, tsarev2018quantum}, quantum simulation \cite{kundu2022quantum}, quantum communication \cite{muralidharan2008perfect,muralidharan2008quantum,sisodia2017teleportation}, quantum imaging \cite{gilaberte2019perspectives} and quantum machine learning \cite{bartkiewicz2020experimental, havlivcek2019supervised, chatterjee2016generalized, schuld2021quantum}. Quantum algorithms are of utmost importance for various tasks, where Shor's factorization algorithm, Grover's search algorithm, Boson sampling, \emph{etc} are structured by exploiting quantum computation and also offer computational speedup \cite{shor1999polynomial, grover1997quantum, aaronson2011computational}. Various platforms have been explored for implementing the algorithms \cite{ladd2010quantum,zagoskin2011quantum,ralph2010optical,georgescu2020trapped,cox2022spin}, among which a quantum optics-based quantum computer considers light as a quantum state for information carrier and exploits the spooky properties of quantum mechanics \cite{ralph2010optical,kok2007linear}. Integrated photonic quantum chips with single photon as information carrier is implemented \cite{wang2020, harris2016} to perform various computational, information communication, and processing tasks \cite{metcalf2014,srikara2020continuous}. On the other hand, the use of another state of light, called a coherent state, is an established candidate for several quantum information means, including quantum teleportation \cite{wang2001,van2001}, quantum key distribution \cite{liao2017}, quantum fingerprinting \cite{arrazola2014,xu2015,kumar2017efficient} and models for universal quantum computer \cite{jeong2002,lund2008}.

In this work, we introduce a quantum optical-based algorithm, where a harmonic oscillator coherent state, being an appropriate model for a laser, is taken as the resource state, $\{|n\rangle\}$:
\begin{equation*}
    \ket{\alpha}=e^{\frac{-|\alpha|^2}{2}}\sum_{n=0}^{\infty}\frac{\alpha^n}{\sqrt{n!}}\ket{n}.
\end{equation*}
Here, $\alpha \in \mathbb{C}$ is a complex coherent state parameter, connected to the average photon number of the laser.

Motivated from the fact that, digital image processing is an emerging real-world field, we prepare our algorithm to perform a quantum image processing (QIP) task. When the resolution of the acquired images is gradually increasing with time, imparting greater control over the decision-making and the information acquisition, it necessitate higher storage \cite{gonzalez2009digital}. Quantum-based hardware is the future solution, which uses quantum correlations and superposition to minimize storage and to maximize computational power. Such integration of storage and processing of images with quantum computation is dealt with by the QIP, which has numerous applications towards
edge detector \cite{yao2017, fan2019, li2020,liu2022, geng2022}, image similarity measurement \cite{yan2012, yan2012next, guanlei2020novel}, image matching \cite{yang2015novel, jiang2016, tezuka2022}, watermarking \cite{luo2022} \emph{etc}. Various models for QIP also exist \cite{su2020new}, some of those also comprise of qubit lattice, flexible representation of the quantum image (FRQI) and novel enhanced quantum representation (NEQR) \cite{venegas, latorre, le, zhang}. We also emphasize that our protocol is \emph{deterministic}, comprising of deterministic gate operations and measurements. This also makes it quite handy for digital image processing algorithms, where the input and the output are nothing but images. While the image retrieval scheme from the output quantum states is challenging \cite{zhang}, we overcome it by using a single resource state, unlike the qubit-based models, where a large number of interconnected qubits are usually necessary. The underlying quantum image representation model exploits point transformation, global intensity transformations and similarity measurement metric weighted by a cosine nonlinear function.

In the next section, we discuss the structure of our algorithm for representing digital image via a multimode coherent state where the architecture is based on linear optical elements. Section \ref{section 3} provides the method of image retrieval and the identification of appropriate coherent state amplitude for optimal retrieval scheme, relying on indistinguishability and quantum uncertainty. The similarity measurement algorithm between a single test image and an image database is described in Sec. \ref{section 4}. Section \ref{section 5} deals with the demonstration of our similarity measurement algorithm for objective quality assessment of images and their distortion-sensitivity, which is followed by conclusions and future outlook in Sec. \ref{section 6}.

\section{Protocol for Encoding the Image into a Multimode Coherent state}
\label{section 2}
An image of $P\cross Q$ resolution is defined by the two-dimensional function $I(x,y)$, where ($x,y$) represents the spatial coordinates of a pixel with intensity $I(x,y)$. For a digital image, each intensity value given by $I(x,y)$ belongs to the range $\mathcal{I}=[0,2^j-1]$, where $j$ is the number of bits per pixel, required to store the intensity labels in terms of a binary string. $j$ is unity for a \emph{binary image}, whereas $j>1$ signifies a \emph{gray-label image}. Our algorithm exploits the representation of the intensity information of an image in terms of the phases of a harmonic oscillator multimode coherent state. For doing so, we start by performing a mapping of the pixel intensity levels ($\mathcal{I}$) to $\theta=[0,\pi/2]$ through an injective function $f$:
\begin{equation}
\label{eq2.1}
    f:s\rightarrow \theta_s=\frac{\pi}{2}\cross\left(\frac{s}{2^j-1}\right),
\end{equation}
where the index $s$ runs from 0 to $2^j-1$, \emph{i.e.} for all intensity labels of $\mathcal{I}$. The relation within the parenthesis of Eq. (\ref{eq2.1}) is basically a min-max normalization process of elements of the range $\mathcal{I}$.
We will provide image representation through a multimode coherent state, which can be generated by chopping a high-intensity coherent field with the use of a \textit{multiport} device.

We define a $T$-mode bosonic system, described by a set of bosonic operators, $\{(\hat{a}_j, \hat{a}_j^\dagger);\textrm{ } j=1,\dots, T\}$, which satisfy the bosonic commutation relation, $[\hat{a}_j,\hat{a}^\dagger_k]=\delta_{jk}$, where $\delta$ is the Kronecker delta function. The transformation relation between the input creation operators, $(a^\dagger_1,\dots,a^\dagger_T)$ and the output creation operators $(b^\dagger_1,\dots,b^\dagger_T)$ are given by,
\begin{align}
    \hat{b}_k^\dagger&=\sum_{j=1}^T u_{kj}\hat{a}_j^\dagger,\textrm{ and }\label{eq2.2}\\
    \hat{a}_j^\dagger&=\sum_{k=1}^T u^*_{jk}\hat{b}_k^\dagger \label{eq2.3},
\end{align}
where, $u_{jk}$ is an element of the discrete $(T\times T)$ unitary matrix, representing the multiport device's transformation. We refer Eqs. (\ref{eq2.2}) and (\ref{eq2.3}), respectively, to designate \emph{forward} and \emph{backward} transformations.

The resource state of the algorithm is a coherent state $|\alpha\rangle$, which is taken as the input to the multiport device and gets transformed into the output mode as
\begin{equation}
    \label{eq2.4}
    \ket{\alpha}_1^{in}\bigotimes_{j=2}^T\ket{0}_j^{in}\rightarrow \bigotimes_{k=1}^T \ket{u^*_{1k}\alpha }_k^{out}.
\end{equation}
    \\

The detailed steps for deriving the above transformation is furnished in Appendix \ref{appendixA} for better understanding.
As we are required to chop the initial coherent state into $T$ identical daughter states under the normalization condition for $u_{1k}$ for conserved photon number, where $u^*_{1k}=1/\sqrt{T}$ for all $k$. Then, the desired input-to-output transformation arrives at
\begin{equation}
\label{eq2.5}                   \ket{\alpha}_1^{in}\bigotimes_{j=2}^T\ket{0}_j^{in}\rightarrow \bigotimes_{k=1}^T\ket{\frac{\alpha}{\sqrt{T}}}_k^{out}.
\end{equation}
Eq. (\ref{eq2.5}) is the tensor-product of the multimode coherent state for image representation. At the circuit label, a multiport device can be implemented via an optical circuit, consisting of a sequence of single and two ports devices, such as phase shifters, beam splitters, and mirrors \cite{reck1994experimental, clements2016optimal}.
\begin{figure*}[t]
\centering
\includegraphics[width=10cm]{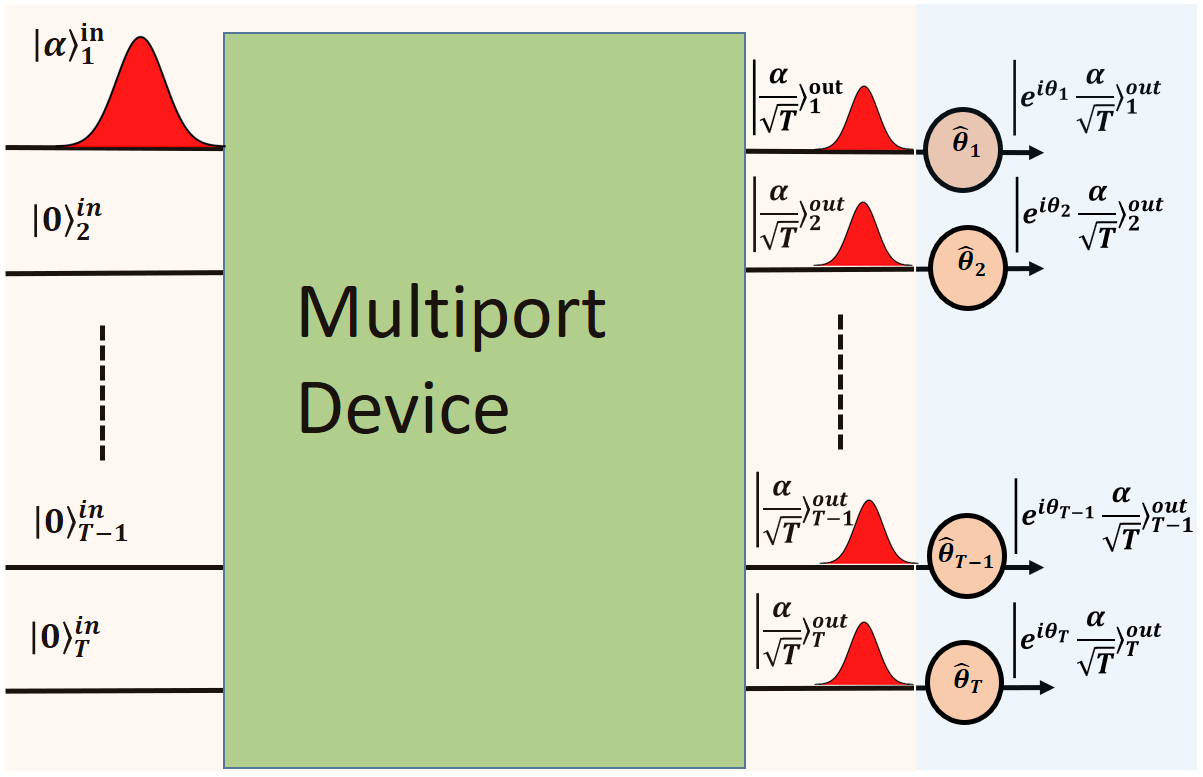}
\caption{Schematic for image representation algorithm. The image representation algorithm consists of two steps: coherent state chopping by a multiport device and pixels' intensity information encoded  through an array of  phase shifters.}
\label{fig1}
\end{figure*}

\begin{figure*}[t]
\centering
\includegraphics[width=10cm]{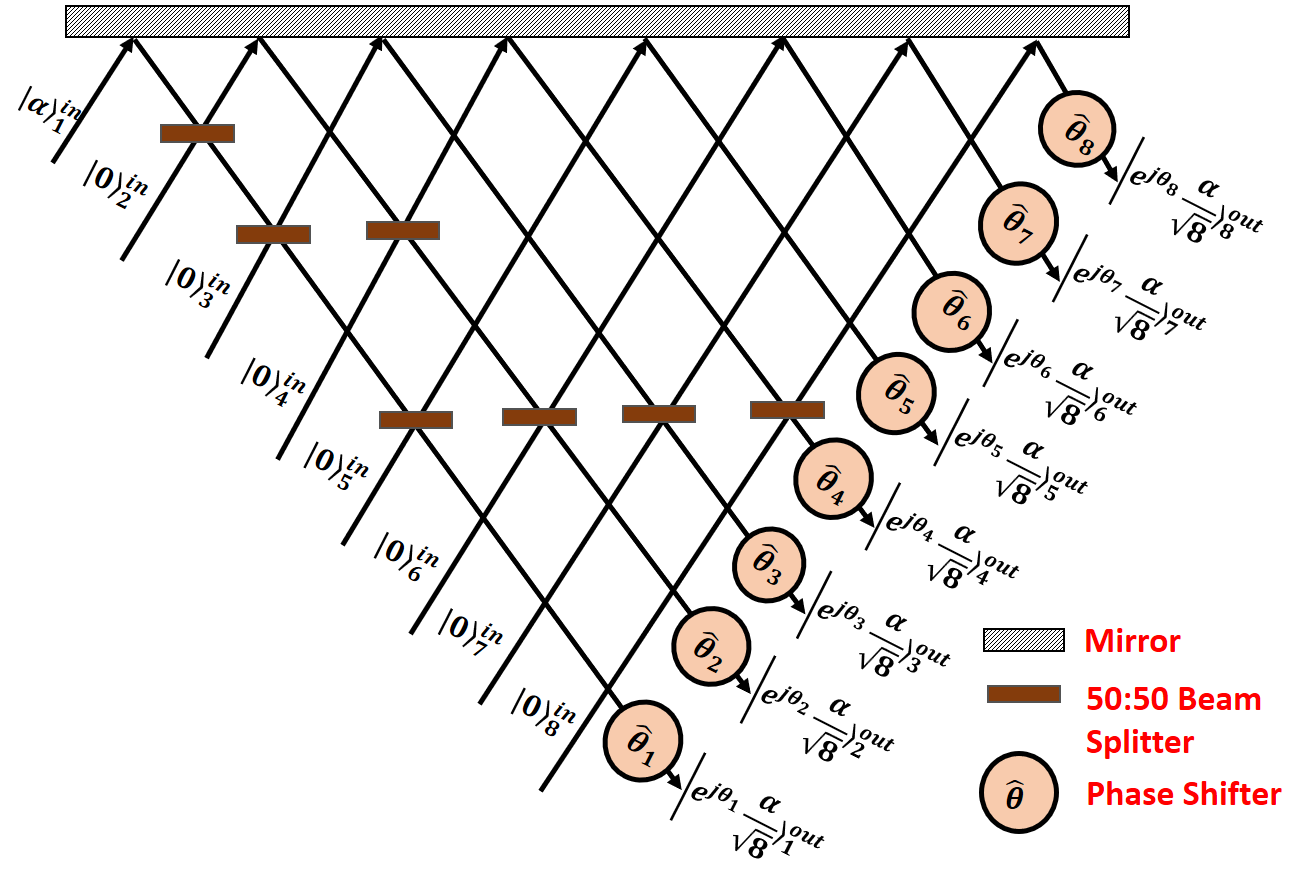}
\caption{Toy example for image representation scheme by using the passive optical network. A combination of beam splitters for chopping a single coherent state into eight daughter coherent states and then encoding intensity values corresponding to pixels' position \textit{via} a parallel array of phase shifters.}
\label{fig2}
\end{figure*}

Now, we describe the steps for encoding the data of an image into the above multimode coherent state. The image data after intensity mapping is represented by a vector of angles, $\Vec{\theta}:=(\theta_1,\theta_2,\dots,\theta_T)$ with dimension $T$ $(=P\cross Q) \in\mathbb{N}$, which is also the total number of pixel of the image under consideration. For image representation, we need to initialize $T$ coherent states, expressed by Eq. (\ref{eq2.5}), where the subscript corresponds to a particular pixel. We call these modes as \textit{pixel modes} because they contain the corresponding intensity information of all individual pixels. The mapping of intensity labels to angles and chopping of the source coherent state are already discussed in Eq. (\ref{eq2.1}) and (\ref{eq2.5}), respectively. Next, we apply the phase shifter, $\hat{\theta}_k=exp(i \theta_k \hat{b}^\dagger_k\hat{b}_k)$, to each pixel mode of the multimode coherent state. Finally, we obtain a state $\ket{I}$, which can represent the input image as a system of the phase-distributed multimode coherent state:
\begin{equation}
\label{eq2.5.1}
    \ket{I}=\bigotimes_{k=1}^{T}\ket{e^{i\theta_{k}}\frac{\alpha}{\sqrt{T}}}_k^{out}.
\end{equation}
Fig. (\ref{fig1}) depicts the steps through which the image can be represented via a system of a multimode coherent state. We also do some intensity transformation processes, like point and global intensity transformations over our  system of image representation (see Appendix \ref{appendixB}). A linear optical network architecture is also provided for state preparation to represent the image. Fig. \ref{fig2} demonstrates this process for an image of $8$-pixels. It  consists of a combination of optical elements, such as $50:50$ beam splitters (BS), phase shifters, and mirrors. The initial part of the optical network consists of a combination of $50:50$ BSs, which are represented in terms of the unitary matrices under the condition, $\hat{a}_1^\dagger=1/\sqrt{T}\sum_{k=1}^{T}\hat{b}_k^\dagger$, \emph{i.e.}, the first initial optical mode is a linear superposition of output modes with uniform weightages. 

Under such superposition, the combination of beam splitters can effectively be used for creating a multimode coherent state. Hence, a diagonal matrix represents the parallel pixels' intensities encoding over the previously prepared multimode coherent state. We are going to evaluate the total unitary transformation matrix followed by required sequence of operations in the proposed optical network.
\vskip .3cm\textbf{Step 1}: We present a generic transformation matrix by considering beam splitters over $T$ optical modes. Say, $BS_{p,q}$ is a $T$-mode matrix, where the beam splitter acts between $p$- and $q$-th modes. Then, a $(T\times T)$ identity matrix undergoes a replacement of the elements ($pp$, $pq$, $qp$ and $qq$) by the elements of a beam splitter, where a matrix for 50:50 BS is known as
\begin{equation}\label{eq2.6}
         BS=\frac{1}{\sqrt{2}}\begin{pmatrix}
                                1 & 1 \\
                                1 & -1
                              \end{pmatrix}.
\end{equation}
Hence, the resultant matrix for $T$ optical modes, where a BS is applied between $p$-$q$ modes, becomes
       \begin{equation}\label{2.7}
         BS_{p,q}=
         \frac{1}{\sqrt{2}}\begin{psmallmatrix}
                                      \sqrt{2} & 0 & \dots &  & \dots & & \dots&0 \\
                                      0 & \sqrt{2} & \dots &  & \dots & & \dots&0 \\
                                      \vdots &  &\ddots  &  &  & & &\vdots   \\
                                       0&\dots  &  & 1 &\dots  & 1&\dots &0 \\
                                       \vdots&  &  &  &\ddots  & & &\vdots \\
                                      0&\dots  &  & 1 & \dots &-1 &\dots &0 \\
                                       \vdots&  &  &  &  & &\ddots &\vdots \\
                                       0&\dots  &  &  &\dots  & & \dots&\sqrt{2}
                                    \end{psmallmatrix}_{(T \times T)}.
       \end{equation}
This treatment is employed to our particular state preparation protocol as depicted in Fig (\ref{fig2}). Initially, the first balanced beam splitter, $BS_{1,2}$, is acting between the $1$st and $2$nd modes. For instance, the corresponding matrix over eight optical modes is written as
      \begin{equation}\label{2.8}
        BS_{1,2}=
        \frac{1}{\sqrt{2}}\begin{psmallmatrix}
                   1 & 1 & 0 & 0 & 0 & 0 & 0 & 0 \\
                   1 & -1 & 0 & 0 & 0 & 0 & 0 & 0 \\
                   0 & 0 & \sqrt{2} & 0 & 0 & 0 & 0 & 0 \\
                   0 & 0 & 0 & \sqrt{2}  & 0 & 0 & 0 & 0 \\
                   0 & 0 & 0 & 0 & \sqrt{2}  & 0 & 0 & 0 \\
                   0 & 0 & 0 & 0 & 0 & \sqrt{2}  & 0 & 0 \\
                   0 & 0 & 0 & 0 & 0 & 0 & \sqrt{2}  & 0 \\
                   0 & 0 & 0 & 0 & 0 & 0 & 0 & \sqrt{2}
                 \end{psmallmatrix}.
      \end{equation}
In the second layer of the optical network, two beam splitters $BS_{1,3}$ and $BS_{2,4}$ act on the optical modes $(1 \textrm{ and } 3)$, and $(2 \textrm{ and } 4)$, respectively. The combined operation is a simple matrix multiplication and is given by
      \begin{equation}\label{2.9}
        (BS_{1,3} BS_{2,4})=
        \frac{1}{\sqrt{2}}\begin{psmallmatrix}
                   1 & \textrm{ }0 & \textrm{ }1 & \textrm{ }0 & \textrm{ }0 & \textrm{ }0 & \textrm{ }0 & \textrm{ }0 \\
                   0 & \textrm{ }1 & \textrm{ }0 & \textrm{ }1 & \textrm{ }0 & \textrm{ }0 & \textrm{ }0 & \textrm{ }0 \\
                   1 & \textrm{ }0 & -1 & \textrm{ }0 & \textrm{ }0 & \textrm{ }0 & \textrm{ }0 & \textrm{ }0 \\
                   0 & \textrm{ }1 & \textrm{ }0 & -1  & \textrm{ }0 & \textrm{ }0 & \textrm{ }0 & \textrm{ }0 \\
                   0 & \textrm{ }0 & \textrm{ }0 & \textrm{ }0 & \sqrt{2}  & \textrm{ }0 & \textrm{ }0 & \textrm{ }0 \\
                   0 & \textrm{ }0 & \textrm{ }0 & \textrm{ }0 & \textrm{ }0 & \sqrt{2}  & \textrm{ }0 & \textrm{ }0 \\
                   0 & \textrm{ }0 & \textrm{ }0 & \textrm{ }0 & \textrm{ }0 & \textrm{ }0 & \sqrt{2}  & \textrm{ }0 \\
                   0 & \textrm{ }0 & \textrm{ }0 & \textrm{ }0 & \textrm{ }0 & \textrm{ }0 & \textrm{ }0 & \sqrt{2}
                 \end{psmallmatrix}.
      \end{equation}
There are four beam splitters in the final layer, acting between optical modes, $(1 \textrm{ and } 5)$, $(2 \textrm{ and } 6)$, $(3 \textrm{ and } 7)$, and $(4 \textrm{ and } 8)$, respectively. The resultant unitary matrix takes the form
      \begin{equation}\label{2.10}
        (BS_{1,5}BS_{2,6}BS_{3,7} BS_{4,8})=
        \frac{1}{\sqrt{2}}\begin{psmallmatrix}
                   1 & \textrm{ }0 & \textrm{ }0 & \textrm{ }0 & \textrm{ }1 & \textrm{ }0 & \textrm{ }0 & \textrm{ }0 \\
                   0 & \textrm{ } 1 & \textrm{ }0 & \textrm{ }0 & \textrm{ }0 & \textrm{ }1 & \textrm{ }0 & \textrm{ }0 \\
                   0 & \textrm{ }0 & \textrm{ }1 & \textrm{ }0 & \textrm{ }0 & \textrm{ }0 & \textrm{ }1 & \textrm{ }0 \\
                   0 & \textrm{ }0 & \textrm{ }0 & \textrm{ }1  & \textrm{ }0 & \textrm{ }0 & \textrm{ }0 & \textrm{ }1 \\
                   1 & \textrm{ }0 & \textrm{ }0 & \textrm{ }0 & -1  & \textrm{ }0 & \textrm{ }0 & \textrm{ }0 \\
                   0 & \textrm{ }1 & \textrm{ }0 & \textrm{ }0 & \textrm{ }0 & -1  & \textrm{ }0 & \textrm{ }0 \\
                   0 & \textrm{ }0 & \textrm{ }1 & \textrm{ }0 & \textrm{ }0 & \textrm{ }0 & -1  & \textrm{ }0 \\
                   0 & \textrm{ }0 & \textrm{ }0 & \textrm{ }1 & \textrm{ }0 & \textrm{ }0 & \textrm{ }0 & -1
                 \end{psmallmatrix}.
      \end{equation}
Hence, the total optical network involving beam splitters can be replaced by the effective unitary operation for an $8$-pixel image as follows.
    \begin{align}\label{2.11}
        U &= (BS_{1,5}Bs_{2,6}BS_{3,7} Bs_{4,8})(BS_{1,3} BS_{2,4})BS_{1,2}  \notag\\
        &=\frac{1}{2\sqrt{2}}\begin{psmallmatrix}
                   1 & \textrm{ }1 & \textrm{ }\sqrt{2} & \textrm{ }0 & \textrm{ }2 & \textrm{ }0 & \textrm{ }0 & \textrm{ }0 \\
                   1 & -1 & \textrm{ }0 & \textrm{ }\sqrt{2} & \textrm{ }0 & \textrm{ }2 & \textrm{ }0 & \textrm{ }0 \\
                   1 & \textrm{ }1 & -\sqrt{2} & \textrm{ }0 & \textrm{ }0 & \textrm{ }0 & \textrm{ }2 & \textrm{ }0 \\
                   1 & -1 & \textrm{ }0 & -\sqrt{2}  & \textrm{ }0 & \textrm{ }0 & \textrm{ }0 & \textrm{ }2 \\
                   1 & \textrm{ }1 & \textrm{ }\sqrt{2} & \textrm{ }0 & -2  & \textrm{ }0 & \textrm{ }0 & \textrm{ }0 \\
                   1 & -1 & \textrm{ }0 &-\sqrt{2} & \textrm{ }0 & -2  & \textrm{ }0 & \textrm{ }0 \\
                   1 & \textrm{ }1 & \textrm{ }\sqrt{2} & \textrm{ }0 & \textrm{ }0 & \textrm{ }0 & -2  & \textrm{ }0 \\
                   1 & -1 & 0 \textrm{ }& -\sqrt{2} & \textrm{ }0 & \textrm{ }0 & \textrm{ }0 & -2
                 \end{psmallmatrix}.
  \end{align}
  It is worth noticing that, one will need seven balanced beam splitters for representing an $8$-pixel image. It can be extrapolated for a $T$-pixel image, where one will require ($T-1$) beam splitters in linear order.
\vskip .3cm
\textbf{Step 2}: The second step is comparatively straightforward with consecutive phase shifters, acting on each mode. The unitary transformation (PS) for a $8$-pixel image is given by
  \begin{align}\label{2.12}
    PS =\bigotimes_{k=1}^8 \hat{\theta}_{k}
      ={\begin{psmallmatrix}
          e^{i\theta_1} & 0 & 0 & 0 & 0 & 0 & 0 & 0 \\
          0 & e^{i\theta_2} & 0 & 0 & 0 & 0 & 0 & 0 \\
          0 & 0 & e^{i\theta_3} & 0 & 0 & 0 & 0 & 0 \\
          0 & 0 & 0 & e^{i\theta_4} & 0 & 0 & 0 & 0 \\
          0 & 0 & 0 & 0 & e^{i\theta_5} & 0 & 0 & 0 \\
          0 & 0 & 0 & 0 & 0 & e^{i\theta_6} & 0 & 0 \\
          0 & 0 & 0 & 0 & 0 & 0 & e^{i\theta_7}& 0 \\
          0 & 0 & 0 & 0 & 0 & 0 & 0 & e^{i\theta_8}
        \end{psmallmatrix}}.
  \end{align}
This optical network can be generalized to a $2^n$ optical mode, where $n$ is a nonzero, positive integer. However, we can also design the protocol where the number of modes is not $2^n$. In that case, we need to modify the constituent beam splitter with a different transmission coefficient, $cos\gamma\textrm{ }( \gamma \in [0,\pi/2])$, such that
\begin{equation}\label{eq2.13}
  BS(\gamma)=\begin{pmatrix}
              cos\gamma  & sin\gamma \\
               sin\gamma & - cos\gamma
             \end{pmatrix},
\end{equation}
keeping the required number of beam splitter same, \emph{i.e.}, ($T-1$) beam splitters for $T$-mode.

\section{Identifying Appropriate Parameter Values for Similarity Measurements through Image Retrieval Scheme}
\label{section 3}
The intensity information of pixels is phase-encoded through the multimode coherent state. Since there is no Hermitian operator defined for phase, hence phase can be estimated but cannot be absolutely measured \cite{pezze2014quantum}. Hence, one usually finds another physical quantity whose measurement over any optical mode of the multimode coherent state is parameterized by the phase of the coherent state. In our case,\emph{ field} intensity  is the physical quantity that serves the purpose. Note that the addition of adjective field before physical observable intensity to distinguish the word intensity label of pixels. For estimating the intensity of the $k$-th pixel, represented by the state, $\ket{e^{i\theta_{k}}\alpha/\sqrt{T}}$, we need to consider an auxiliary coherent state, $\ket{e^{i\theta_{r}}\alpha/\sqrt{T}}$. Both are fed into a four-port balanced BS, where the working of a BS is mentioned in Eq. \ref{eq2.6}. We exploit the quantum state transformation in a BS with the input coherent states, $\ket{\alpha}$ and $\ket{\beta}$ :
\begin{align}
\label{eq3.1}
    \ket{\alpha}\ket{\beta}\rightarrow\ket{\frac{\alpha + \beta}{\sqrt{2}}}\ket{\frac{\alpha - \beta}{\sqrt{2}}}.
\end{align}
For the intensity retrieval of $k$-th pixel, we replace the input beams, $\ket{\alpha}$ and $\ket{\beta}$  by $\ket{e^{i\theta_{k}}\alpha/\sqrt{T}}$  and $\ket{e^{i\theta_{r}}\alpha/\sqrt{T}}$, respectively. Then, Eq. (\ref{eq3.1}) transforms as
\begin{multline}
\label{eq3.2}
    \ket{e^{i\theta_{k}}\frac{\alpha}{\sqrt{T}}}\ket{e^{i\theta_{r}}\frac{\alpha}{\sqrt{T}}}\rightarrow \\ \ket{(e^{i\theta_{k}}+e^{i\theta_{r}})\frac{\alpha}{\sqrt{2T}}}\ket{(e^{i\theta_{k}}-e^{i\theta_{r}})\frac{\alpha}{\sqrt{2T}}}.
\end{multline}
Transformed coherent states contain the angle-encoded intensity labels by measuring either of these modes. Before that, we see how field intensity measurement is related to the expectation value of the number operator for any light state. A single-mode  electromagnetic field of amplitude $E_0$, frequency $\omega_k$, and wave vector $\mathbf{\kappa}$ can be described by
 \begin{equation}
     \hat{E}_k(\textbf{r},t)=E_0\left[\hat{a}_{k}e^{i(\mathbf{\kappa}.\textbf{r}-\omega_kt)} + \hat{a}_k^\dagger e^{-i(\mathbf{\kappa}.\textbf{r}-\omega_kt)}\right].
 \end{equation}
Hence, the field intensity of the coherent field under consideration is represented by the average photon number: $I_k(\textbf{r},t) \propto \langle\hat{a}_k^\dagger\hat{a}_k\rangle $ or $\langle\hat{n}_k\rangle$. The expectation value of this number operator over the second output mode, $\ket{(e^{i\theta_{k}}-e^{i\theta_{r}})\alpha/\sqrt{2T}}$, of Eq. (\ref{eq3.2}) becomes
\begin{equation}
\label{eq3.4}
    \langle\hat{n}_k\rangle=\frac{|\alpha|^2}{T}\left(1-cos(\theta_k-\theta_r)\right),
\end{equation}
which is dependent on the relative phase between the pixel mode and the auxiliary mode, where $0\leq(\theta_k-\theta_r)\leq \pi/2$. The phase of the auxiliary signal can also be fixed to zero and then Eq. (\ref{eq3.4}) reduces to
\begin{equation}
\label{eq3.5}
    \langle\hat{n}_k\rangle=\frac{|\alpha|^2}{T}\left(1-cos(\theta_k)\right).
\end{equation}
The above equation can be used to retrieve the intensity of the $k$-th pixel. The intensity for other pixels can be obtained by following the same procedure. Note that, we need to run this algorithm once, simultaneously for all the pixel, to retrieve the data for the image, which makes the image retrieval scheme deterministic in nature. However, this deterministic nature of the algorithm requires two \textit{distinguishable} coherent states to represent two consecutive intensity labels,  $\theta_{s+1}=(\pi/2)\left[(s+1)/(2^j-1)\right]$ and $\theta_{s}=(\pi/2)\left[(s)/(2^j-1)\right]$, following Eq. (\ref{eq2.1}). The measurement \textit{uncertainity} also plays a role in addition.

\begin{figure}[ht]
    \centering
    \includegraphics[width=7.5cm]{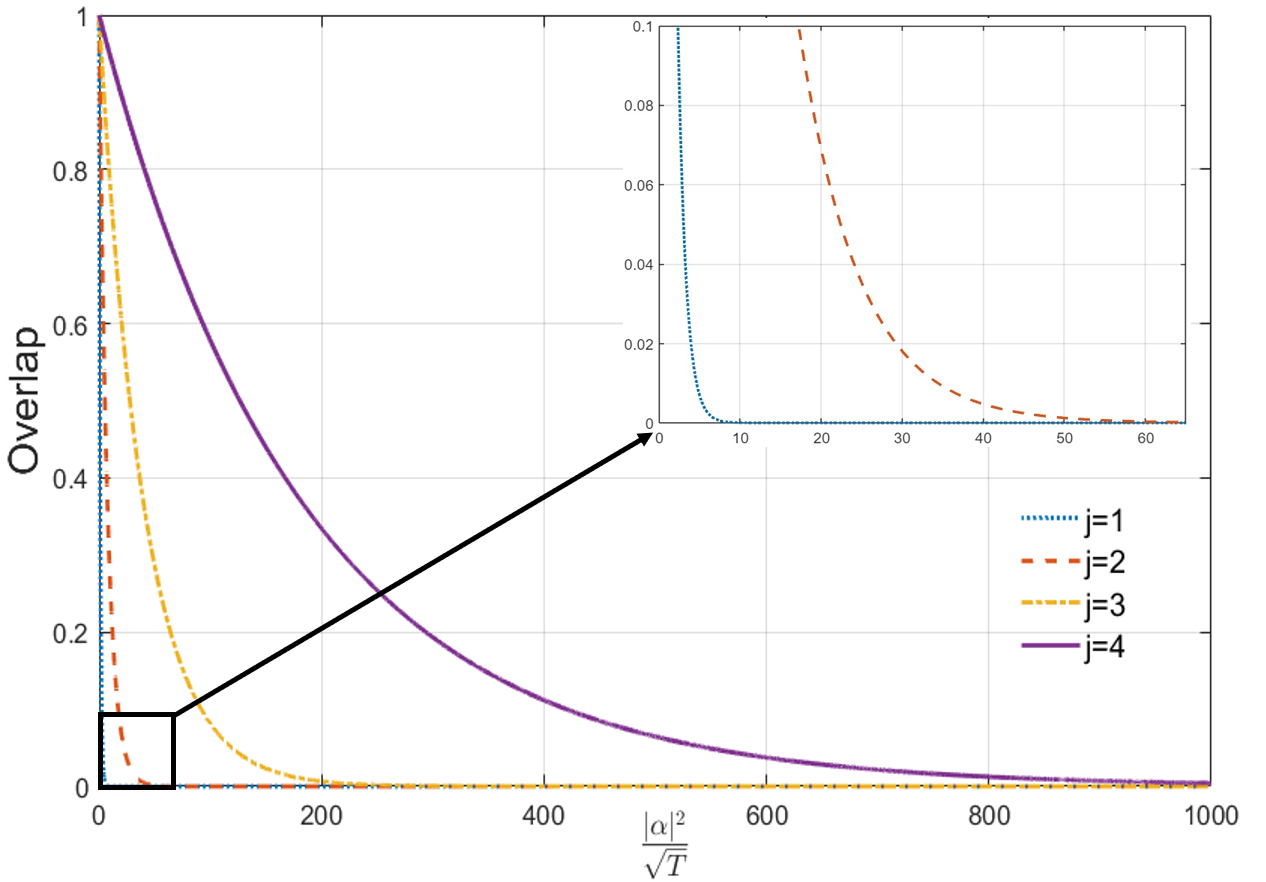}
    \caption{Overlap plot as a function of the coherent state amplitude to find the optimal coherent state amplitude for different image types, which have intensity range, $[0,2^j-1]$, such that, two coherent states, encoding two consecutive intensity labels for particular image type, must be distinguishable.}
    \label{fig 3}
\end{figure}

The distinguishability of two coherent states, $\ket{\alpha}$ and $\ket{\beta}$, depends on their \textit{overlap}  $|\braket{\alpha}{\beta}|^2=exp(-|\alpha-\beta|^2)$. The two coherent states become orthogonal, \emph{i.e.}, $|\braket{\alpha}{\beta}|^2\approx 0$, when their coherent state parameters take the fairly distinct value and they can be distinguished in a field  intensity measurement device. If we consider two input coherent states, $\ket{e^{i\theta_{s+1}}\alpha/\sqrt{T}}$ and $\ket{e^{i\theta_{s}}\alpha/\sqrt{T}}$, which upon going through a beam splitter operation, turn into  $\ket{(e^{i\theta_{s+1}}-e^{i\theta_{r}})\alpha/\sqrt{2T}}$ and $\ket{(e^{i\theta_s}-e^{i\theta_{r}})\alpha/\sqrt{2T}}$, respectively, where $|\theta_{s+1}-\theta_s|=\pi/[2(2^j-1)]$ as per Eq. (\ref{eq3.2}). The corresponding overlap becomes
\begin{equation*}
    exp\left\{\frac{-|\alpha|^2}{T}\left[1-cos\left(\frac{\pi}{2(2^j-1)}\right)\right]\right\}.
\end{equation*}
The distinguishability can be decided from the overlap plot as depicted in Fig.( \ref{fig 3}). A proper choice of the optimal
amplitude for enabling the coherent states towards encoding can be performed. For instance, we find from Fig. (\ref{fig 3})  that, the optimal coherent state amplitude for an image with $1$-bit ($2$-bits) per pixel is obtained as $\alpha/\sqrt{T}\approx$ $\sqrt{2.3}$ ($\sqrt{17.2}$), when the overlap between the successive intensity levels tends to $10\%$.

\begin{figure*}[ht]
    \includegraphics[width=17cm]{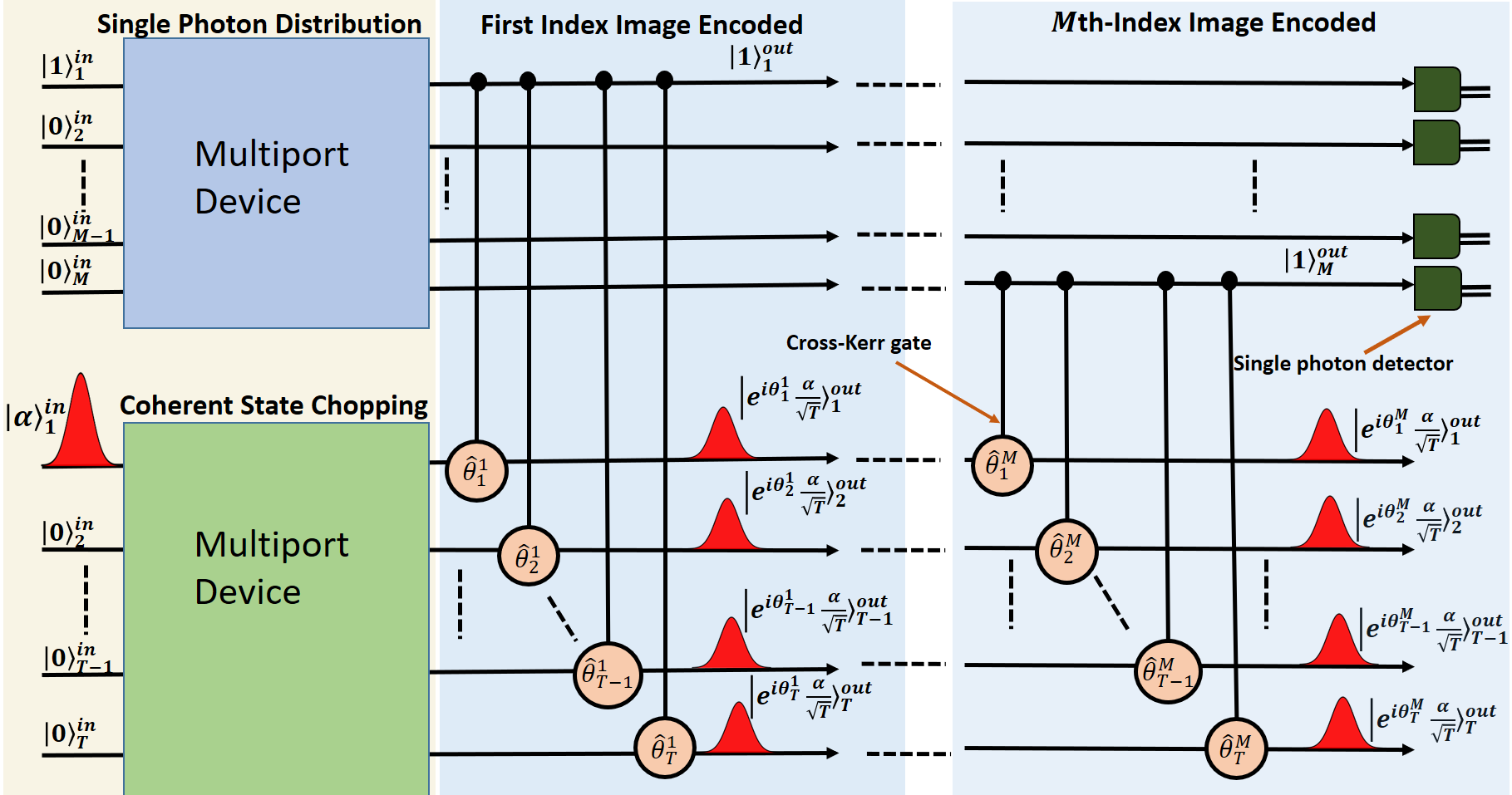}
    \caption{Schematic for an optical quantum circuit for image database representation. The first step consists of two multiport devices, which respectively distribute a single photon and a chopped coherent state in the required number of optical modes. Then, we use $M\cross T$ sequences of cross-Kerr nonlinear gates to embed the database information. In a single run, anyone among $M$ single photon detectors detects a photon from the index modes. This detection implies the same index image from the image database, information of which is encoded in pixel modes.}
    \label{fig 4}
\end{figure*}

\begin{figure*}[ht]
    \includegraphics[width=18cm,height=10cm]{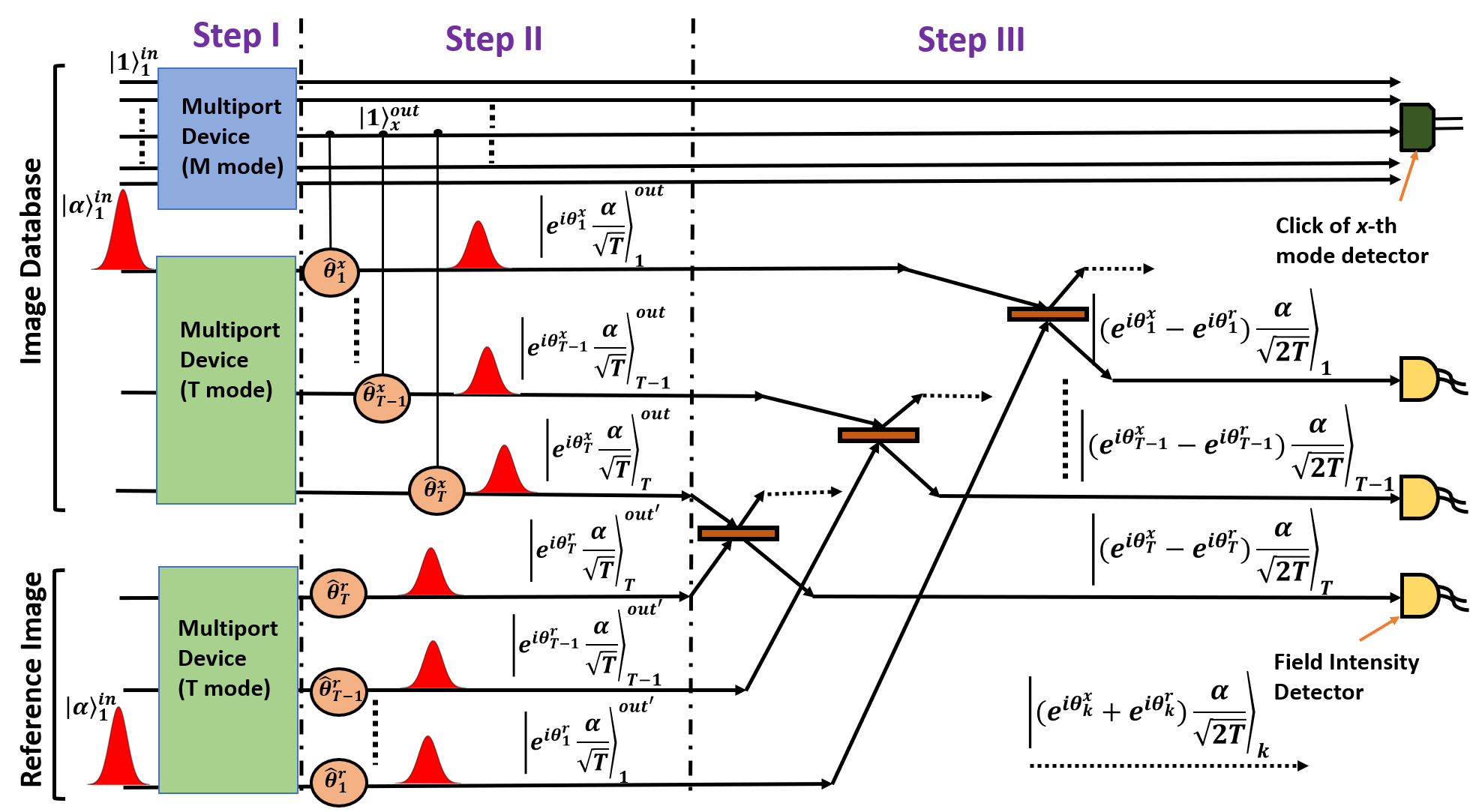}
    \caption{Schematic of all the steps, I, II, and III, demarcated by the vertical dot-dash lines. Step-I shows the input state for reference image and the image database. The inverse enumeration of the qmodes of the multiport device is to make the representation clean. Step-II represents encoding scheme, which is following by the similarity measurement protocol and detections in step-III.}
    \label{fig 4a}
\end{figure*}
In addition to fulfilling the condition of distinguishability for the multimode coherent state, we also compute the associated uncertainty in the measurement of field intensities of these coherent states, which is proportional to the square root of the average value of photon numbers of the coherent states:
\begin{align}
\label{Eq.3.5}
    \Delta \hat{n}_k=\sqrt{\langle{\hat{n}_k\rangle}}.
\end{align}
Such fluctuation is known as \textit{shot noise} \cite{ataman2018phase}. For correctly \emph{estimating} the value of $k$-th pixel's intensity label, we need to add the shot noise to the average value of the photon number. Therefore, the resultant intensity of the $k$-th pixel can be written as
\begin{align}
   \langle\hat{n}_k\rangle \pm \Delta \hat{n}_k  &= \langle\hat{n}_k\rangle \pm \sqrt{\langle{\hat{n}_k\rangle}} \notag \\
   &= \frac{|\alpha|^2}{T}\left(1-cos(\theta_k)\right) \pm \left(\frac{|\alpha|^2}{T}\left(1-cos(\theta_k)\right)\right)^{\frac{1}{2}}.
\end{align}
Table \ref{table 1} lists the field intensity measurements including shot noise for the given optimized values of coherent state amplitudes. Two different image types are considered. From column-4 of Tab. \ref{table 1}, we find that our optimized amplitude for the particular image type differs from the estimated field intensity values for different pixel-label. Hence, the actual value of the intensity label of the pixel can be retrieved from the measured value with the knowledge of the shot noise.

\begin{table}[t]
    \begin{tabular}{|c|c|c|c|c|}
    \hline
     Image & Optimal   & Intensity & Measured Field\\
     Type & Coherent & Labels & Intensity\\
      & Amplitude &   & Value \\
     & $\alpha/\sqrt{T}$ & $\theta_s$ & $\langle\hat{n}_k\rangle \pm \Delta \hat{n}_k$ \\ \hline\hline
     \multirow{2}{*}{1-bit per pixel } & \multirow{2}{*}{$\sqrt{2.3}$}& 0 & $ 0.0 \pm 0.0 $ \\ \cline{3-4}
      & & $\pi/2$ & $ 2.3 \pm 1.516 $ \\ \hline \hline
     \multirow{4}{*}{2-bits per pixel } & \multirow{4}{*}{$\sqrt{17.2}$}& 0 &  $ 0.0 \pm 0.0 $ \\  \cline{3-4}
     & & $\pi/6$ & $ 2.304 \pm 1.518 $ \\\cline{3-4}
     & & $\pi/3$ & $ 8.599 \pm 2.932 $ \\\cline{3-4}
     & & $\pi/2$ & $ 17.2 \pm 4.147 $ \\ \hline
    \end{tabular}

    \caption{\label{table 1} Estimating field intensities for different intensity labels at the optimized coherent amplitudes for two types of images. Here, the auxiliary label is taken as zero ($\theta_r=0$).}
\end{table}

\section{Similarity Measurement Protocol}
\label{section 4}

\subsection{Similarity between Two Images:}
\label{subsec4.1}
Consider  $I(x,y)$ and $I'(x,y)$ are two images of the same resolution, $P\times Q$. By using the above method of image preparation, the states of the two images are
\begin{equation}
    \ket{I}=\bigotimes_{k=1}^T \ket{e^{i\theta_k}\frac{\alpha}{\sqrt{T}}}_k^{out} \\ \quad \textrm{and}\quad
     \ket{I'}=\bigotimes_{k=1}^T \ket{e^{i\theta'_k}\frac{\alpha}{\sqrt{T}}}_k^{out}.
\end{equation}
An array of $T$-balanced beam splitters are used to mix the corresponding modes of both images. The outputs of each beam splitter are expressed by the relation in Eq. (\ref{eq3.2}). These are followed by $T$ field intensity measuring devices to measure the field  intensity of the second output mode of each beam splitter. Such field intensity measurement is given by Eq. (\ref{eq3.4}), which for all $T$ intensity measurements yields the similarity metric between two images:
\begin{equation}
\label{eq4.1}
    sim(I,I')=\frac{1}{T}\sum_{k=1}^T cos(\theta_k-\theta'_k)=\left(1-\frac{1}{|\alpha|^2}\sum_{k=1}^T\langle\hat{n}_k\rangle\right).
\end{equation}
This computes the pixel-to-pixel similarity between two images, which is weighted by the cosine function. Therefore, we refer to this metric as the \emph{cosine-similarity metric}.

\begin{figure*}[t]
    \centering
    \includegraphics[width=16cm]{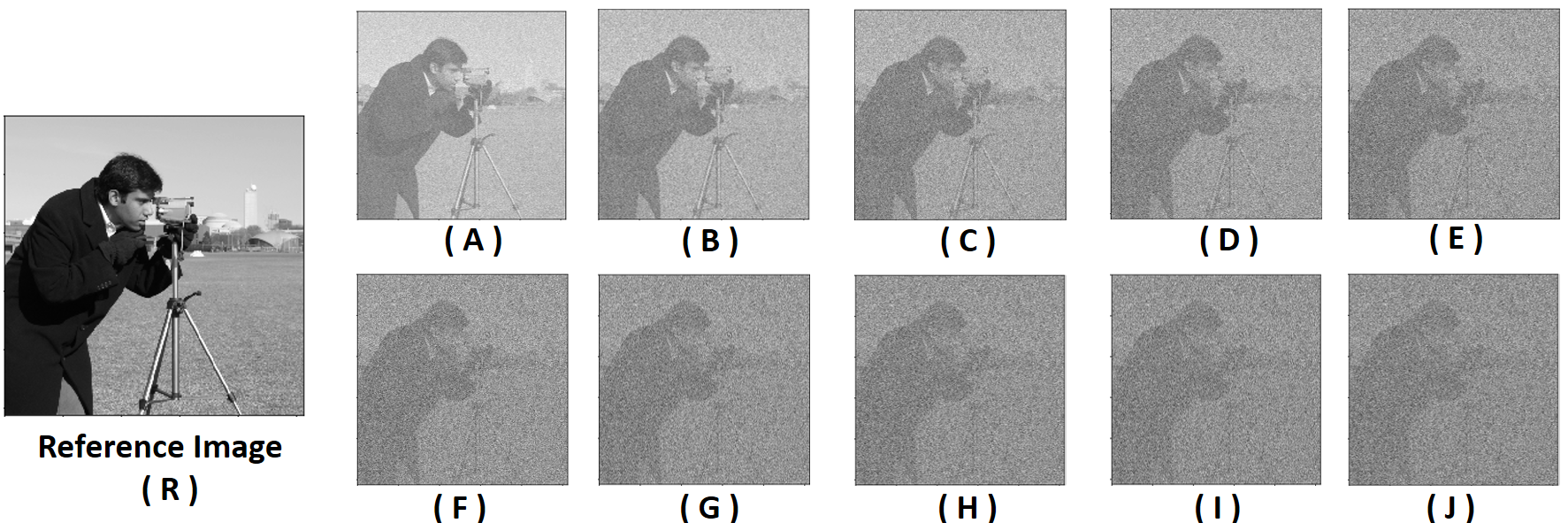}
    \caption{Reference image (R) along with the cluster of distorted images (A-J), which are created by adding Gaussian noise in a layer-wise manner, \textit{i.e.}, succeeding image is created by adding next layer of Gaussian random noise over the preceding image. Note that all layers of Gaussian random noises have the same parameters.}
    \label{fig 6}
\end{figure*}

\subsection{Similarity between an Image Database and a Reference Image:}
\label{subsec4.2}

The working scheme of the similarity measurement between an image database and a reference image relies on \emph{quantum} resources, such as single photon source, series of cross-Kerr gates and single photon detectors \cite{hadfield2009single}. We consider a database of $M$ images, each of $P\times Q$ resolution: $\mathcal{D}=\{I^m\}_{m=1}^M$, which is represented as
\begin{equation}
\label{eq4.2}
    \ket{\mathcal{D}}=\frac{1}{\sqrt{M}}\sum_{m=1}^M\ket{1}_m^{out}\bigotimes_{k=1}^T\ket {e^{i\theta_k^m}\frac{\alpha}{\sqrt{T}}}_k^{out},
\end{equation}
where $\frac{1}{\sqrt{M}}\sum_{m=1}^M\ket{1}_{b_m}$ denotes the quantum state in which a single photon is equally probable in $M$ optical modes. The index, $m$, represents the entry of an image in the image database and is called the \textit{index mode}. The multiport device used for the image database encoding is identical to the setup in Sec. \ref{section 2}. The schematic along with the optical architecture is delineated in Fig. (\ref{fig 4}). Then, the cross-Kerr nonlinear interactions are used to encode the intensity information of images over a multimode coherent state using a single photon state as a controlled state. The phase shift can be produced in the target mode (say mode $2$) by the cross-Kerr interaction, depending on the presence of photons in the control mode (say mode $1$). The Hamiltonian of the cross-Kerr interaction is $\hat{H}_{Kerr}=-\chi\hat{n}_1\hat{n}_2$, where $\chi$ is the strength of the nonlinearity and $\hat{n}_1$ $(\hat{n}_2)$ is the photon number operator of the control (target) mode. State transformation due to cross-Kerr interaction through the evolution operator, $exp(-i\hat{H}_{Kerr}t)$, becomes
\begin{equation}
 \label{eq4.3}   \ket{1}_1\ket{\alpha}_2\rightarrow\ket{1}_1\ket{e^{i\theta}\alpha}_2,
\end{equation}
where $\theta=\chi t$ \cite{nielsen2002quantum,jeong2006quantum, he2011cross}. It is very difficult to achieve the typical requirement of the phase shift, as natural Kerr media offer inadequate nonlinearities ($\chi \approx 10^{-18}$). However, this can be further enhanced $\chi $ to higher order, through electromagnetically induced transparency (EIT) \cite{schmidt1996giant,fushman2008controlled,kang2003observation, lo2011electromagnetically, hoi2013giant, venkataraman2013phase,tiarks2016optical}. Particularly, Daniel Tiarks \textit{et al.} have experimentally demonstrated a $\pi$-phase shift in the first photon due to the presence of a second single photon pulse \cite{tiarks2016optical}. The image encoding of the reference image is expressed in Eq.(\ref{eq2.5.1}). In the following, we describe detailed working of the scheme which is depicted in Fig. \ref{fig 4a}. The vertical dot-dash lines are to distinguish the steps I, II and III. \\
\textbf{Step I}: A single photon state is considered as the first mode of the multiport device having total $M$ modes. Moreover, two multimode coherent states emerge from two additional multiport devices of $T$ modes. Hence, the initial state consists of three quantum registers:
 \begin{equation}
     \label{eq4.4}
     \frac{1}{\sqrt{M}}\sum_{m=1}^M\ket{1}_m^{out}\bigotimes_{k=1}^T\ket {\frac{\alpha}{\sqrt{T}}}_k^{out}\bigotimes_{k=1}^T\ket {\frac{\alpha}{\sqrt{T}}}_{k}^{out'}.
 \end{equation} \\
\textbf{Step II}: Assuming that the $x$-th single photon detector has a click, the corresponding $T$ cross-Kerr gates will get activated to encode the $x$-th from the database. Simultaneously, a series of phase shifters embed the reference image into the second multimode coherent state. The resultant output state becomes
\begin{equation}
    \label{eq4.5}
    \ket{1}_x^{out}\bigotimes_{k=1}^T\ket {e^{i\theta_k^x}\frac{\alpha}{\sqrt{T}}}_k^{out}\bigotimes_{k=1}^T\ket {e^{i\theta_k^r}\frac{\alpha}{\sqrt{T}}}_{k}^{out'},
\end{equation}\\
where indices, $x$ and $r$, stand for the $x$-th image and the reference image, respectively.\\
\textbf{Step III}: The final step is already briefed in subsection (\ref{subsec4.1}), where we compute the similarity between the $x$-th indexed image and the reference image. The expression of the similarity metric can be rewritten as
\begin{equation}
    \label{eq4.6}
     \frac{1}{T}\sum_{k=1}^T cos(\theta_k^x-\theta_k^r)=\left(1-\frac{1}{|\alpha|^2}\sum_{k=1}^T\langle\hat{n}_k\rangle\right),
\end{equation}
which is the estimate of the output of the field intensity measurements over the required quantum modes (see Eq.(\ref{eq3.4})). A single run of the algorithm provides the similarity between a member (say $x$-th image), belonging to the image database, and a reference image, whereas a comparative study of all the similarity values will require $\sim\mathcal{O}(M)$ runs of the algorithm. Each time, the similarity measurement between the $x$-th image and the reference image is guided by Eq. (\ref{eq4.6}).

\section{A Demonstration of the Algorithm: Objective Quality Assessment of Images}
\label{section 5}

\begin{table*}
\centering
    \begin{tabular}{|c|c|c|c|}
    \hline
    Sl. No.& Image Pair & Cosine-Similarity & Mean Square Error\\
    & & Metric & Metric\\
    \hline \hline
    1  & R-R & 1 & 0  \\
    \hline
    2 & R-A & 0.978 & 0.044 \\ \hline
    3 & R-B & 0.955 & 0.091 \\ \hline
    4 & R-C & 0.941 & 0.121 \\ \hline
    5 & R-D & 0.931 & 0.142 \\ \hline
    6 & R-E & 0.924 & 0.157 \\ \hline
    7 & R-F & 0.918 & 0.168 \\ \hline
    8 & R-G & 0.914 & 0.176 \\ \hline
    9 & R-H & 0.911 & 0.183 \\ \hline
    10 & R-I & 0.908 & 0.189 \\ \hline
    11 & R-J & 0.906 & 0.194 \\ \hline
    \end{tabular}
\caption{Quantitative estimate of the two metrics: cosine-similarity metric (column-II) and mean square error metric (column-III), between the original image (R) and a set of distorted images (A-J).}
\label{Table 2}
\end{table*}

Here, we demonstrate the working of the algorithm for assessing the quality of the images belonging to an image database. A known image database is taken, so that the output of the algorithm can be verified by observation. We prepare the image database by gradually disturbing the image quality in sequence. Usually, Image quality can be distorted \emph{via} a number of processes, such as lossy compression artifacts, image acquisition, blurring artifacts, and due to various types of noises \emph{etc} \cite{wang2006modern}. To quantify the level of distortion in a given image, one needs a suitable technique, which comes under the \textit{image quality assessment} (IQA) algorithm \cite{athar2019comprehensive, okarma2021combined}. Such algorithms are broadly categorized into two classes: subjective quality assessment and objective quality assessment. The subjective quality assessment, being a human-based assessment, is time-consuming, tedious, and expensive. On the other hand, objective quality assessment is an algorithm, based on different mathematical assertions. Here, we take the primitive objective quality assessment algorithm, which takes the mean square error (MSE) metric to measure the dissimilarity between the distorted and non-distorted images \cite{wan2009mean}. This \emph{full-reference} metric determines the closeness between two digital images by exploiting the difference in the statistical distributions of pixel values. Suppose, $I(x,y)$ and $\Tilde{I}(x,y)$ denote the original and distorted images, respectively. Then, the MSE metric is given by
\begin{equation}
    MSE=\frac{1}{PQ}\sum_{x=1}^{P}\sum_{y=1}^{Q}\left[I(x,y)-\Tilde{I}(x,y)\right]^2.
\end{equation}
Here, we discuss the feasibility of our metric as expressed in Eq. (\ref{eq4.2}) and compare its result with a well-known MSE metric for measuring the quality of distorted images. Let us outline the main information for preparing the cluster of distorted images. The pixel intensity value of the reference image varies from $[0,255]$ to $[0,1]$ by using the min-max normalization method. The first distorted image is created by adding Gaussian random noise to the reference image \cite{lone2018noise}. Other distorted images in the database are also created following the same procedure, \emph{i.e.}, by adding Gaussian random noise with the same parameters to the preceding image. Once this cluster is prepared, we consider these images as unknown entries. In Fig. \ref{fig 6}, we have displayed the prepared database of ten such distorted images (A-J) with their reference image (R) for illustration. One can also observe the gradual reduction of similarity from (A) to (J) with respect to the reference image (R). To verify our algorithm, the same ranking should be obtained from the measurements. We compute the level of their closeness with respect to the original image by using two metrics: cosine-similarity metric and MSE metric. It is important to specify that, each distorted image is also passed through the min-max normalization step and then mapped their pixels in terms of angles in the range, $[0,\frac{\pi}{2}]$, using a mapping function, stated in Eq. (\ref{eq2.1}). Both, cosine-similarity metric and MSE metric, for all the images are quantified in Tab. \ref{Table 2}. As expected, the cosine-similarity and MSE for the image pair (R-R) have come out as one and zero, respectively. The cosine-similarity reveals the decreasing values as we proceed from image pair (R-A) to image pair (R-J), which becomes consistent to our observation too. On the contrary, the computed MSE gradually increases from image pair (R-A) to image pair (R-J), which is also expected for gradually worsened images. Hence, both the measures, the cosine-similarity metric, and the MSE metric, are quite consistent to each other and our algorithm is shown to be quite useful for objective quality assessment.

\begin{figure}[h]
    \centering
    \includegraphics[width=8cm,height=6cm]{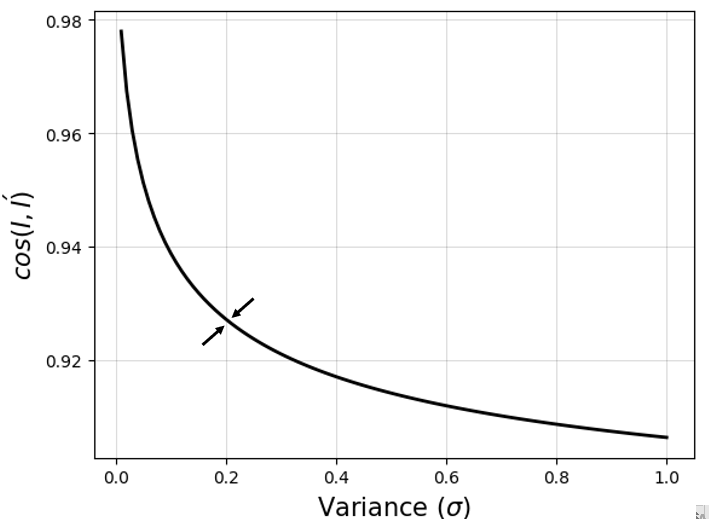}
    \caption{The similarity value upon changing the variance from $0.01$ to $1$ of the Gaussian random noise. The arrow mark indicates the point, $\sigma=0.2$, for which a separate study is carried out to find the distortion sensitivity of the algorithm. Here, the mean value of the Gaussian random noise is taken as zero.}
    \label{fig 7}
\end{figure}

\subsection{Image Distortion Sensitivity of the Algorithm:} So far, we have examined the sufficient merit of the algorithm in distinguishing quality of images. We would like to further see the efficiency of the protocol against various levels of noises. In Fig. \ref{fig 7}, the cosine-similarity metric is evaluated at the different levels of distortion added to the reference image, where each point on the curve corresponds to a test image. As expected and observed in the previous demonstration, Fig. \ref{fig 7} implies that an image gradually looses its similarity with the reference image when the noise variance becomes gradually greater. Hence, the algorithm will provide the correct similarity ranking for all level of noises. In addition, an observation of the slope (similarity/noise ratio) at the low noise level (steep slope) implies that our algorithm is highly sensitive to test the low noise level distortion. For high noise level, the image becomes almost washed out by noise and finding similarity in that case will be naturally less effective.
\begin{table}
 \centering
    \begin{tabular}{|P{25mm}|P{35mm}|c|}
    \hline
    Noise Level ($\sigma_i$)  & $cos(I,I')  $ & Ranking \\
    \hline\hline
      $0.2+0.0$   & $0.9271967133928533$ & $1$ \\ \hline
      $0.2+1\times 10^{-14}$   & $0.9271967133928524$ & $2$ \\ \hline
      $0.2+2\times 10^{-14}$   & $0.9271967133928517$ & $3$ \\ \hline
      $0.2+3\times 10^{-14}$   & $0.9271967133928511$ & $4$ \\ \hline
      $0.2+4\times 10^{-14}$   & $0.9271967133928504$ & $5$ \\ \hline
      $0.2+5\times 10^{-14}$   & $0.9271967133928494$ & $6$ \\ \hline
      $0.2+6\times 10^{-14}$   & $0.9271967133928484$ & $7$ \\ \hline
      $0.2+7\times 10^{-14}$   & $0.9271967133928477$ & $8$ \\ \hline
      $0.2+8\times 10^{-14}$   & $0.9271967133928471$ & $9$ \\ \hline
      $0.2+9\times 10^{-14}$   & $0.9271967133928463$ & $10$ \\ \hline
      $0.2+1\times 10^{-13}$   & $0.9271967133928454$ & $11$ \\ \hline
    \end{tabular}
    \caption{The noise level, having zero mean and variance $\sigma$ is analyzed around $\sigma=0.2$ from Fig.(\ref{fig 7}). Similarity value and ranking are given in columns II and III, corresponding to each infinitesimal change between two successive distorted images: $\Delta \sigma=\sigma_i-\sigma_{i-1}=10^{-14}$ and $\sigma_i=\sigma_0+\Delta\sigma \times i$, where $i=1,\dots,10$ and $\sigma_0=0.2.$}
    \label{Table 3}
\end{table}

We also emphasize on checking the robustness of cosine-similarity metric against an infinitesimal amount of noise added between the successive images in the database. We have chosen an arbitrary point from the curve in Fig. \ref{fig 7}, which is indicated by the arrow marks around $\sigma=0.2$, to carry out the distortion sensitivity of the algorithm. The result is provided in the  Tab. \ref{Table 3} for checking whether the algorithm offers correct similarity ranking between the reference image ($I$) and distorted image ($I'$), even if an infinitesimal amount of distortion is present.
Column-I reflects the amount of distortion: $\Delta \sigma=\sigma_i-\sigma_{i-1}=10^{-14}$ and $\sigma_i=\sigma_0+\Delta\sigma \times i$, where $i=1,\dots,10$ and $\sigma_0=0.2.$, whereas the corresponding similarity value and ranking are given in columns II and III. The result shows that, we are able to obtain a correct ranking even for the negligible distortion ($10^{-14}$), which further strengthen the applicability of the algorithm. We could not go beyond $10^{-14}$,  \emph{i.e.}, $\Delta \sigma < 10^{-14}$, in the noise level due to exceeding our machine precision for that similarity value. However, it is possible in principle. The result is not restricted by the initial noise level ($\sigma=0.2$) and can be checked around any other noise level.
\\
\\

In addition, objective quality assessment of images is only taken as an example to show the merit of our algorithm and is never limited to such applications. Furthermore, the underlying protocol has the potential to identify the similarity of any database with its reference data other than images.

\section{Conclusion}
\label{section 6}
We have reported an algorithm, where an image is represented as a phase-distributed multimode coherent state. The optical architecture for state preparation is also provided by taking a toy example and its working through mathematical operations. The reported scheme exploits only a single coherent state as a constituent, along with a linear number of optical elements, \emph{i.e.}, $(T-1)$ and $T$ numbers of beam splitters and phase shifters, respectively. The resolution of the images is set to $T$ while keeping the depth of the optical circuit exponentially small $log_2T$. The provided retrieval scheme to get the optimal field intensity is deterministic in nature and also minimizes the requirement for running the retrieval circuit, which is in contrast to the qubit-based models where multiple runs are usually required. Upon successful execution of the algorithm, we achieve the similarity measurement protocol through the cosine-similarity metric and the mean square error metric. We quantify the input and output parameters for images for verifying the reported algorithm. The demonstration is performed for a prepared noise-masked image database to identify the similarity with an unmasked image. We are also able to verify our result for the negligible distortion ($10^{-14}$), which further strengthen the applicability of the algorithm. The obtained result is applied to the objective quality assessment technique of distorted images. The algorithm can be easily extended to a supervised machine learning task, where the similarity between labeled instances and test instances is evaluated, as one does during the training stage of the nearest-neighbours algorithm.

\section{Appendix}
\subsection{Transformation of  Multimode Coherent State}
\label{appendixA}
We present a generic transformation of the multimode coherent state guided by a linear optical circuit and then connect it to the state transformation as discussed in Eq. (\ref{eq2.4}). The input multimode coherent state is represented by
\begin{equation}
\label{eqA1}
    \bigotimes_{j=1}^T\ket{\alpha_j}_j^{in}=e^{\left(-\sum_{j=1}^T\frac{|\alpha_j|^2}{2}\right)}exp\left(-\sum_{j=1}^T\alpha_j\hat{a}_j^\dagger\right)\bigotimes_{j=1}^T\ket{0}_j^{in},
\end{equation}
which is fed into the multiport device. Using backward relation (Eq. (\ref{eq2.3})), the input state transforms into
\begin{align}
\label{eqA2}
    &\rightarrow e^{\left(-\sum_{j=1}^T\frac{|\alpha_j|^2}{2}\right)}exp\left(-\sum_{j=1}^T\alpha_j\sum_{k=1}^Tu^*_{jk}\hat{b}^\dagger_k\right)\bigotimes_{k=1}^T\ket{0}_k^{out}\notag \\
    &=e^{\left(-\sum_{j=1}^T\frac{|\alpha_j|^2}{2}\right)}exp \left(-\sum_{k=1}^T(\sum_{j=1}^Tu^*_{jk}\alpha_j)\hat{b}^\dagger_k\right)\bigotimes_{k=1}^T\ket{0}_k^{out}.
\end{align}
The typical constraint for the unitary matrix, $\sum_{j=1}^T|u^*_{jk}|^2=1\forall \textrm{ }k$, which also arises in the discrete Fourier transform matrix and the Hadamard transform matrix, modifies the term: $e^{\left(-\sum_{j=1}^T\frac{|\alpha_j|^2}{2}\right)}=e^{\left(-\sum_{j=1}^T\frac{|u^*_{jk}\alpha_j|^2}{2}\right)}$, and transforms the state into
\begin{align}
    &=e^{\left(-\sum_{j=1}^T\frac{|u^*_{jk}\alpha_j|^2}{2}\right)}exp \left(-\sum_{k=1}^T(\sum_{j=1}^Tu^*_{jk}\alpha_j)\hat{b}^\dagger_k\right)\bigotimes_{k=1}^T\ket{0}_l^{out}\notag \\
    &=\bigotimes_{k=1}^T\ket{\sum_{j=1}^Tu^*_{jk}\alpha_j}_k^{out}.\label{eqA3}
\end{align}

According to the above outcome, the input state, $\ket{1}_1^{in}\otimes_{j=2}^T\ket{0}_j^{in}$, will result an output state, $\bigotimes_{k=1}^T \ket{u^*_{1k}\alpha }_k^{out}$. This can be inferred from Eq. (\ref{eq2.4}), by putting $\alpha_j=0\textrm{ }\forall j,\textrm{ except }j=1$.


\subsection{Image Representation through Intensity Transformations:}
\label{appendixB}
Transformation in a quantum computer is a unitary evolution, $\ket{I'}=U \ket{I}$, between the input and the output intensities, $\ket{I}$ and $\ket{I'}$, respectively. Below, we mention the point- and the global-intensity transformations. It is worth reiterating that, the pixel intensity levels (the transformation of which is discussed here) were mapped to gradually modulated phases of the coherent state in Eq. (\ref{eq2.1}).

\textit{Point Intensity Transformation}: Point transformation changes the individual pixel-intensity value. Suppose we want to change the intensity of the \textit{k}-th pixel from $\ket{e^{i\theta_k}\;\alpha/\sqrt{T}}$ to $\ket{e^{i(\theta_k+\theta^{'}_k)}\;\alpha/\sqrt{T}}$ by applying a phase shifter, $exp(i\theta_k^{'}\hat{a}_k^\dagger\hat{a}_k)$, on the \textit{k}-th optical mode. Then, the pixel mode transforms as
\begin{align}
    \ket{e^{i\theta_k}\frac{\alpha}{\sqrt{T}}}_k^{in} &= exp(i\theta_k^{'}\hat{a}_k^\dagger\hat{a}_k)\ket{e^{i\theta_k}\frac{\alpha}{\sqrt{T}}}_k^{in} \notag \\
    &\rightarrow\ket{e^{i(\theta_k+\theta^{'}_k)}\frac{\alpha}{\sqrt{T}}}_k^{out}.
\end{align}

\textit{Global Intensity Transformation}: In this case, we simultaneously perform the transformation for all the pixels. It can be done by applying the single phase shifter, $exp(i\theta^{'}\hat{a}^\dagger\hat{a})$, just before the multiport device. When this phase shifted input transforms \emph{via} the multiport device, we get the final state,
\begin{equation}
    \ket{e^{i\theta^{'}}\alpha}_1^{in}\bigotimes_{j=2}^T\ket{0}_j^{in}\rightarrow \bigotimes_{k=1}^T\ket{e^{i\theta^{'}}\frac{\alpha}{\sqrt{T}}}_k^{out},
\end{equation}
as mentioned in Eq. (\ref{eq2.5}). This transformation is followed by the protocol in Sec. \ref{section 2}.

\bibliography{bib}

\end{document}